\documentclass[twocolumn,showpacs,prb,amsmath,amssymb,superscriptaddress]{revtex4}

\usepackage{graphicx}
\usepackage{bm}
\usepackage{bbm}
\usepackage{mathbbol}
\newcommand{\identity}{\mathbbm{1}}

\DeclareMathOperator{\Max}{Max}

\DeclareMathAlphabet{\mathpzc}{OT1}{pzc}{m}{it}

\begin{document}

\title{Decoherence induced by anisotropic hyperfine interaction in Si
  spin qubits}
\author{W. M. \surname{Witzel}} 
\affiliation{Condensed Matter Theory Center,
Department of Physics, University of Maryland, College Park, MD 20742-4111} 
\author{Xuedong \surname{Hu}}
\affiliation{Department of Physics, University at Buffalo, SUNY, Buffalo, NY
14260-1500} 
\author{S. \surname{Das Sarma}}
\affiliation{Condensed Matter Theory Center,
Department of Physics, University of Maryland, College Park, MD 20742-4111} 
\date{\today}
\begin{abstract}
We study Si:P donor electron spin decoherence due to anisotropic
hyperfine (AHF) interaction with the surrounding nuclear spin bath.  
In particular, we clarify the electron spin echo envelope modulation 
(ESEEM) in the Si:P system and the resonance-like contributions from
nuclear spins in various shells away from the P atoms.  We suggest an
approach to minimize AHF-induced decoherence by avoiding the resonances 
and orienting an applied magnetic field along directions that can 
periodically eliminate contributions from the dominant nearest 
neighbor atoms. 
Our remarkable agreement with experiment demonstrates nearly complete 
understanding of electron spin decoherence in Si:P when
combining ESEEM, spectral diffusion, instantaneous diffusion, and
spin-lattice relaxation.
\end{abstract}


\maketitle

\section{Introduction}

The prospect of scalable solid state quantum information processing has 
created extensive interest in the study of coherent manipulation of single 
electron spins confined in semiconductor quantum dots or donor states.  
The clarification of single electron spin properties such as decoherence and
quantum control should also be valuable scientifically in the field of 
nanostructure physics and very useful technologically in the context of
spintronics and spin quantum memory.
%
%

%
%
%
Among the many semiconductor host materials, silicon is particularly 
enticing\cite{DasSarma} due to its deep-rooted connection to the modern
microelectronics industry.  Furthermore, Si provides a remarkably quiet 
environment for electron\cite{Vrijen,Friesen} and nuclear\cite{Kane} 
spin qubits since $^{28}$Si, its most abundant isotope, has no net 
nuclear spin, and spin-orbit interaction is weak in Si.  
In this context, Si, with P donor electron spins 
as qubits, is an ideal semiconductor material for quantum information
processing.  The donor electron spin dephasing time $T_2$ as
measured directly (and calculated theoretically) by spin echo decay is
extremely long (many milliseconds in contrast to microseconds in
GaAs) and can be further enhanced via isotopic purification.  
However, a strong anisotropic hyperfine (AHF) interaction in Si
%
%
presents a formidable challenge for Si spin quantum computation;
it produces finite spin dephasing relatively quickly that cannot be 
eliminated by spin refocusing techniques, and could potentially
nullify the advantage of long spin $T_2$ time in Si.  

This paper details the effects of the AHF interaction on donor
electron spin qubits in Si:P.  We discuss this interaction and present
our Hamiltonian formalism in Sec.~\ref{AHF_interactions}.  In
Secs.~\ref{FID} and \ref{ESEEM}, we formulate the problem of electron
spin evolution and decoherence due to the AHF interaction in the context
of free induction decay and (Hahn) spin echoes, respectively.
We compare theoretical computations with experimentally observed spin 
echo envelopes in Sec.~\ref{vsExperiment}, where
we demonstrate nearly complete theoretical understanding of the 
decoherence of the Si:P electron spin qubit. 
Combining AHF-induced electron spin echo envelope modulation (ESEEM) 
with decoherence from
spectral diffusion, instantaneous diffusion, and spin-lattice relaxation, we
account for all major sources of decoherence and achieve nearly perfect
quantitative agreement with Hahn echo decay experiments.  We believe that
AHF interaction supplies the final piece of the puzzle with respect
to understanding Si:P electron spin decoherence.
Furthermore, Sec.~\ref{suppressingESEEM}
provides valuable insight into AHF-induced decoherence and a
prescription for its suppression.
Without proper treatment, this decoherence can
violate the stringent fault-tolerant requirements of qubit fidelity
much sooner than the nominal $T_2$ time suggests.
While there are {\it no} existing spin echo techniques to generally
remove AHF-induced echo modulations,\cite{TyryshkinP} we 
suggest a concrete method for suppressing this decoherence and clarify 
requirements on the applied magnetic field for spin quantum computation 
in Si.  We give concluding remarks in Sec.~\ref{conclusion}.

\section{Anisotropic Hyperfine Interaction}
\label{AHF_interactions}

The hyperfine (HF) interaction between an electron and a nuclear spin
describes the magnetic dipolar coupling between the two spin 
species.\cite{Schweiger}  With $\hat{\bf S}$ denoting the spin operator
of the electron and $\hat{\bf I}$ that of the nucleus, 
the HF Hamiltonian is given by
$\hat{\cal H}_{HF} =  \hat{\bm I} \cdot \bm{A} \cdot \hat{\bm S}$,
where the hyperfine tensor ${\bf A}$ is given by
\begin{equation}
\label{HFtensor}
\bm{A}_{ij} = \gamma_I \gamma_S \hbar^2 \left(\frac{8 \pi}{3} 
\left\lvert \Psi(\bm{0}) \right\rvert^2 \delta_{ij} + 
\left\langle \Psi \left\vert \frac{3 x_i x_j - \bm{r}^2 \delta_{ij}}{\bm{r}^5}
\right\vert \Psi \right\rangle \right),
\end{equation}
with the electron position measured relative to the site of the nucleus, 
$\Psi$ the electron wavefunction, and $\gamma_S$ and
$\gamma_I$ the gyromagnetic ratios of the electron
and the nucleus.  The first term of
Eq.~(\ref{HFtensor}) is the {\it isotropic} Fermi contact HF interaction.
%
%
The second term is {\it anisotropic}.  Which part of the interaction is
more important depends on the electron wave function.  For example, the GaAs
conduction band minimum occurs at the $\Gamma$-point of the Brillouin zone, 
where the electron Bloch function is mostly atomic s-type, so that HF
interaction in GaAs between an electron near the conduction band minimum and
the surrounding nuclear spins is essentially isotropic.  On the other hand,
the degenerate conduction band minimum for Si occurs close to the X-point of 
the Brillouin zone, where the electron Bloch functions have significant 
contributions from p- and d-atomic-orbitals,\cite{Ivey75,Jancu,Koiller} 
so that HF interaction between an electron near the conduction band minimum, 
such as an electron confined to a donor
or a quantum dot, and the surrounding nuclear spins has strong anisotropic
characteristics.  Indeed, AHF interaction has been studied extensively in the 
Si:P system in the 1960s and 1970s.\cite{Feher, Hale69, Ivey75}  The strength
of AHF has been accurately measured\cite{Hale69} and calculated\cite{Ivey75}
for the phosphorus donor electron.
In the context of solid state spin quantum computation, however, much of the
existing literature only take into account the contact HF
[first term in Eq.~(\ref{HFtensor})] in considering
electron spin decoherence.

To analyze how AHF leads to spin decoherence, we consider a single P donor 
in Si, with the donor-bound electron interacting
with the P and randomly distributed $^{29}$Si nuclear spins.
We assume the limit of a strong magnetic field ($> 100~\mbox{mT}$ is
sufficient) applied in the $z$
direction such that electron spin flips are suppressed due to its large
Zeeman energy.  Since $\gamma_S \gg \gamma_I$, it is appropriate to
take the limit where $\hat{S}_z$ is conserved but not $\hat{I}_z$ (of any
nucleus).  In this limit we write the Hamiltonian (in $\hbar = 1$ unit) 
as $\hat{\cal H} = \hat{\cal H}_0 +  \sum_n \hat{\cal H}_n$ with
\begin{eqnarray}
\label{H0}
\hat{\cal H}_0 &=& \omega_S \hat{S}_z + A_P \hat{S}_z \hat{I}_{z}^{P} - 
\omega_P \hat{I}_{z}^{P},
\\
\label{Hn}
\hat{\cal H}_n &=& A_n \hat{S}_z \hat{I}_{nz} + B_n \hat{S}_z
\hat{I}_{nx'} 
- \omega_I \hat{I}_{nz}.
\end{eqnarray}
We separate the Hamiltonian into $\hat{\cal H}_0$, involving the
electron Zeeman energy and the donor nucleus, and $\hat{\cal H}_n$, involving
the $n$th $^{29}$Si nucleus in the surrounding lattice (other Si
isotopes have zero spin).  In our notation, $\hat{\bm S}$, $\hat{\bm I}^{P}$, 
and $\hat{\bm I}_{n}$ denote spin operators of the electron, P nucleus, and 
the $n$th $^{29}$Si nucleus, respectively.  $\hat{I}_{nx'}$ gives the nuclear 
spin operator in an $x'$-axis orientation so that there is no $\hat{S}_z
\hat{I}_{ny'}$ contribution (thus we generally have a different $x'$ 
orientation for each $n$).  Given an applied magnetic field strength of $B$,
we define $\omega_{\Box} = \gamma_{\Box} B$ as the Zeeman frequency
for the electron, P nucleus, or a $^{29}$Si nucleus with 
$\Box=S, P,~\mbox{or}~I$ respectively.
$A_P$ denotes the HF coupling between the electron and the P nucleus.
Both contact HF and the $\hat{S}_z \hat{I}_z$ part of the AHF 
interaction are contained in $A_n$.  The remaining AHF interaction in our 
strong field limit is contained in $B_n$ and gives the
relevant anisotropy, mixing different directional components of
$\hat{\bm S}$ and $\hat{\bm I}$.

Qualitatively, due to the anisotropic term $B_n \hat{S}_z \hat{I}_{nx'}$ in 
$\hat{\cal H}_n$, the quantization axis for the precession of 
the $^{29}$Si nuclear spin is dependent on the state of the electron
spin.  Conversely, the electron spin is affected by the precession
of the nuclear spin.
The resulting electron spin free induction decay (FID) in Si:P 
has been explored in Ref.~[\onlinecite{Saikin03}] and will be briefly reviewed
in Sec.~\ref{FID}.  It is shown\cite{Saikin03} that the donor electron spin
could 
lose more than 1\% of its coherence after {\it only} about 10 $\mu$s if a
$^{29}$Si atom is in one of the nearest neighbor (E-shell) sites.
This would be disastrous for quantum computation where the error rate
must stay below typical fault tolerance requirements of 
$10^{-6} - 10^{-4}$.  Fortunately, as we will show
in Sec.~\ref{suppressingESEEM}, AHF-induced
decoherence may be drastically suppressed by applying precisely timed
pulses and a sufficiently strong magnetic field along special
directions of high symmetry.

\section{Effect of Anisotropic Hyperfine on Free Induction Decay}
\label{FID}

$\hat{\cal H}_n$ of Eq.~(\ref{Hn}) for different $n$ commute with each 
other so that the free evolution operator, in the interaction picture, 
becomes the product of evolution operators for each $n$:
\begin{equation}
\label{freeEvolProduct}
\hat{U}_0(t) = e^{-i \hat{\cal H} t} = e^{-i \hat{\cal H}_0 t} 
\prod_n e^{-i \hat{\cal H}_n t}.
\end{equation}
We may write this evolution operator in the form
\begin{equation}
\label{freeEvolDiagForm}
\hat{U}_0(t) = \hat{U}_0^{+}(t) \hat{P}_{\uparrow} + \hat{U}_0^{-}(t) 
\hat{P}_{\downarrow}
\end{equation}
where $\hat{P}_{\uparrow} = \lvert \uparrow \rangle \langle \uparrow \rvert$
and $\hat{P}_{\downarrow} = \lvert \downarrow \rangle \langle \downarrow
\rvert$ are up and down projection operators for the electron spin
In other words, $\hat{U}_0^{\pm}(t)$ denotes the free evolution of the
nuclei given an electron spin that is either up or down. 
The free evolution operator may take the form of Eq.~(\ref{freeEvolDiagForm})
because the Hamiltonian only involves $\hat{S}_z$ and no other electron spin
operator.
For the same reason, only dephasing occurs for the electron spin and thus 
only the off-diagonal
elements of the electron spin's density matrix may evolve.  Assuming
unpolarized (random) nuclei, after time $t$ the off-diagonal element
of the electron spin's density matrix becomes\cite{Saikin03}
\begin{equation}
\label{FreeInduction}
\frac{\langle \downarrow \vert \rho(t) \vert \uparrow \rangle}{\langle 
\downarrow \vert \rho(0) \vert \uparrow \rangle}
= \prod_n \left\{ a_n^2 \cos{\left(\frac{\Delta \omega_n}{2} t\right)}
+ b_n^2 \cos{\left(\bar{\omega}_{n} t\right)}\right\},
\end{equation}
where $\Delta \omega_n = \omega_{n+} - \omega_{n-}$, $\bar{\omega}_n =
(\omega_{n+} + \omega_{n-})/2$, with
\begin{equation}
\label{omega_n}
\omega_{n\pm} = \sqrt{\left(\pm\frac{A_n}{2} - \omega_I\right)^2 +
  \left(\frac{B_n}{2}\right)^2}, 
\end{equation}
and $a_n^2 + b_n^2 = 1$ (details in Ref.~[\onlinecite{Saikin03}]).
$\omega_{n\pm}$ are precession frequencies for nucleus $n$.
Nucleus $n$ will precess at a frequency of
$\omega_{n+}$ or $\omega_{n-}$ given an up or down
electron spin, respectively.  In general, if the electron spin is in a 
superposition of up and down states, the nuclear spin dynamics will
contain both these frequencies.  As mentioned before, numerical evaluation of 
Eq.~(\ref{FreeInduction}) shows\cite{Saikin03} that the donor electron spin 
could lose more than 1\% of its coherence in $\sim$ 10 $\mu$s if a $^{29}$Si 
atom is in one of the nearest neighbor (E-shell) sites.

\section{Electron Spin Echo Envelope Modulations}
\label{ESEEM}

The key question now is whether the AHF-induced electron spin decoherence
can be suppressed.  
It is well known that spin echo techniques such as Hahn echo can be used to
remove dephasing caused by the spatial variation of local magnetic
fields (the inhomogeneous broadening). 
However, the AHF-induced FID is a dynamical effect and, as
such, cannot be removed by Hahn echo.  Instead, AHF causes the echo envelope 
to oscillate, which is known within the electron spin resonance community
as ESEEM.\cite{Schweiger,Mims72,Dikanov}  This effect is particularly useful 
in chemistry for the identification of nuclear spin species of 
molecular sites \cite{Schweiger,Dikanov}.  
The focus of the present study is to investigate ESEEM in the Si:P 
system\cite{Tyryshkin60ms,TyryshkinSiCoherence,Abe,Ferretti} and explore possible ways to 
reduce the decoherence effect of AHF interaction with
$^{29}$Si in the context of spin quantum computation.

%

\begin{figure}
\includegraphics[width=3in]{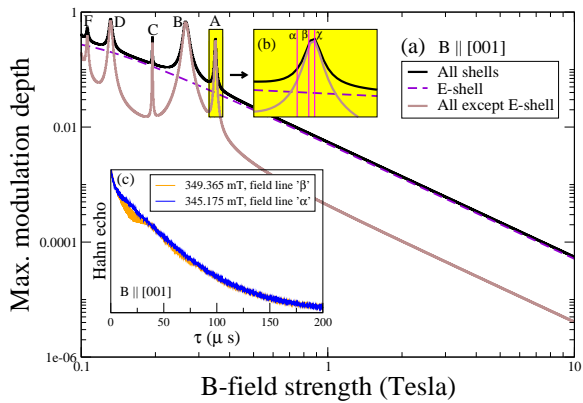}
\caption{
(Color online)
(a) Maximum modulation depth [Eq.~(\ref{maxDepth})] 
in natural Si averaged over isotopic configurations
with an applied magnetic field, $B$, parallel to the
$[001]$ lattice direction considering
all shells (provided in Ref.~[\onlinecite{Ivey75}]),
just E-shell sites (nearest neighbors of the P
donor), and all shells except the E-shell.  
This maximum depth gives the worst-case scenario of constructively
interfering nuclear ESEEM contributions and is useful for identifying
the major sources of modulations.
When one shell of nuclei dominate, the worst-case
maximum modulation depth is close to the actual observed modulation depth.
Near the cancellation condition,
$\omega_I \sim A_n/2$, for each shell of nuclei is a peak labeled by the
shell letter where that shell dominates; away from these peaks, the
E-shell is seen to dominate.  
%
%
(b) Enlargement of the A-shell peak marking three
field strengths, $\alpha$, $\beta$, and $\chi$,
used by experiments presented in this Article.
(c) Corresponding experimental\cite{TyryshkinP} Hahn echo decay at 
field strengths $\alpha$ and $\beta$ for the same Si:P sample.
%
%
Relatively high doping, $10^{16}$ P/cm$^3$, results in fast
exponential relaxation due to instantaneous diffusion, 
but ESEEM is still observed.
%
%
\label{FigModDepth}
}
\end{figure}

%
%
%
%
%
In the Hahn echo sequence, after initializing the electron spin 
(which is necessary for experimental decoherence measurements,
but is neither applicable to qubit preservation nor necessary in the 
current discussion), the system evolves freely
for a time $\tau$, then a resonant pulse rotates the spin by $\pi$ about an
axis perpendicular to the magnetic field, then an echo is observed after
another $\tau$ in time (for shorthand, we use
$\tau \rightarrow \pi \rightarrow \tau$ to denote this sequence).
For convenience in our formalism, though it makes no difference, we
apply a second $\pi$ pulse in order to bring the spin back to its
original orientation (apart from decoherence): $\tau \rightarrow \pi
\rightarrow \tau \rightarrow \pi$.  
Assuming an ideal applied $\pi$ pulse and free evolution in the
diagonal form of Eq.~(\ref{freeEvolDiagForm}), the evolution over the
echo sequence has the diagonal form
\begin{eqnarray}
\label{HahnEvol}
\hat{U}_{\mbox{\tiny Hahn}} = 
\hat{U}_{\mbox{\tiny Hahn}}^{+} \hat{P}_{\uparrow} +
\hat{U}_{\mbox{\tiny Hahn}}^{-} \hat{P}_{\downarrow}, \\
\label{HahnEvolUpDown}
\hat{U}_{\mbox{\tiny Hahn}}^{\pm} = \hat{U}_{0}^{\mp}(\tau)
\hat{U}_{0}^{\pm}(\tau).
\end{eqnarray}

Mathematically, spin echo is the magnitude of the expectation value of
the electron spin (a measure of the fidelity of the qubit) after the quantum
system evolves according to the echo sequence [Eq.~(\ref{HahnEvol})].
As it is physically measured, the echo is the average spin of an ensemble of
electron spins that undergo the Hahn echo evolution.  Assuming only
$\hat{S}_z$ electron spin interactions, as we do, maximum decoherence occurs
for an electron that is initially perpendicular to the $z$ direction (the
direction of the applied magnetic field), and the spin echo is more precisely
defined with such an initial electron spin state.  In the experimental
measurement, the ensemble of electron spins are initialized to point in the
same direction perpendicular to this axis.  Where $\langle...\rangle$ is the
quantum mechanical average over initial bath states, this Hahn echo envelope
may be expressed as\cite{witzelSD}
\begin{equation}
\label{v_Hahn}
V(\tau)=
\left\langle
\left[\hat{U}_{\mbox{\tiny Hahn}}^{-}\right]^{\dag} 
\hat{U}_{\mbox{\tiny Hahn}}^{+} \right\rangle.
\end{equation}
In computing the expectation value, we average over all initial nuclear
states, assuming a completely disordered and unpolarized initial bath as in
Sec.~\ref{FID}.

Electron spin echo decays as a function of $\tau$ due to internuclear
interactions.  In the current study we do not focus on this process, known as
spectral diffusion, which has been studied by two of us\cite{witzelSD}
previously.  On top of this decay, AHF coupling [Eq.~(\ref{Hn})] produces
envelope modulations or ESEEM.\cite{Schweiger,Dikanov}  Neglecting the
internuclear interactions, the ESEEM due to each nucleus factors into the
Hahn echo evolution separately [Eq.~(\ref{freeEvolProduct})].
%
%
Given that only a fraction $f$ of the Si nuclei have nonzero spin
($^{29}$Si), we have the following ESEEM amplitude averaged with respect to
isotope configurations:\cite{Mims72, Reijerse91}
\begin{eqnarray}
\label{echo_mod}
V(\tau) & = & \prod_n \left[(1 - f) + f V_n(\tau) \right], \\
%
%
%
%
\label{echo_n}
V_n(\tau) &=& 1 - \frac{k_n}{2}\left(1 - \cos{(\omega_{n+} \tau)}\right)
\left(1 - \cos{(\omega_{n-} \tau)}\right),~~~~ \\
k_n &=& \left(\omega_I B_n\right)^2 / \left(\omega_{n+}
  \omega_{n-}\right)^2,
\end{eqnarray}
where $\omega_{n\pm}$ are nuclear precessional frequencies defined by
Eq.~(\ref{omega_n}).  In the literature, $k_n$ is called the modulation depth 
parameter.\cite{Reijerse91}  The maximum modulation (deviation from 1) of
$V_n(\tau)$ is $2 k_n$, so that $k_n$ is a measure of modulation amplitude.  
In the ``worst-case'' scenario, when modulations from all nuclei combine
constructively, the maximum possible modulation depth averaged over isotopic
configurations is given by
\begin{equation}
\label{maxDepth}
\Max{\left(1 - V(\tau)\right)} = 1 - \prod_n \left[1 - 2 f k_n\right].
\end{equation}
In Fig.~\ref{FigModDepth}~(a), we show this maximum modulation as a function
of field strength due to various nuclear shells (symmetry-related sets of
lattice sites\cite{Ivey75}).  We use experimentally determined contact and
AHF coupling constants for $22$ nuclear shells (which include about $150$
symmetry-related nuclear sites) taken from Ref.~[\onlinecite{Ivey75}] and
Ref.~[\onlinecite{Hale69}].

\section{Understanding Experimental Observation of Spin Echo Decay and
Modulation}
\label{vsExperiment}

The Hahn echo decay of Si:P donor electron spin has been studied in several
recent experiments.\cite{Tyryshkin60ms,TyryshkinSiCoherence,Abe,Ferretti}  A previous quantum theory
of nuclear-induced 
spectral diffusion (SD) by two of us\cite{witzelSD} shows very good
quantitative and qualitative agreement with these experimentally determined
decay curves.  However, the agreement is not perfect, especially at short
times, because we previously neglected the effects of AHF interactions that
produce modulations of the echo envelope, i.e., ESEEM.  We now add this last
piece to the puzzle to acheive truly remarkable agreement with experimental
echo decay curves and thus demonstrate a nearly complete understanding of
Si:P donor electron spin decoherence.

\begin{figure}
\includegraphics[width=3in]{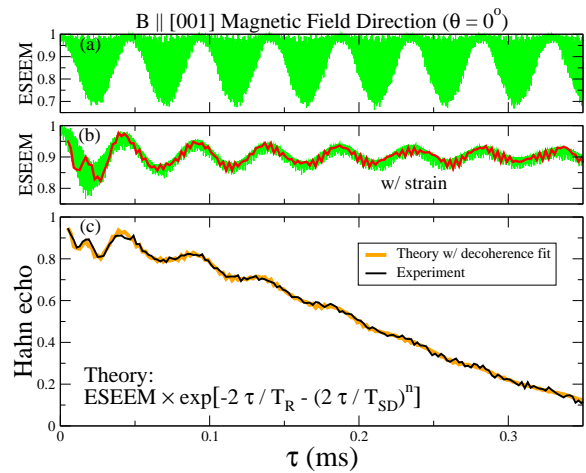}
\caption{
(Color online)
AHF-induced ESEEM in Si:P with an applied magnetic field in the [001]
  direction. (a) Pure AHF-induced ESEEM for a single electron
  spin.  The green ``blob'' is one curve with high frequency
  components. 
(b) Before matching the ESEEM to experiment, we must account
 for strain effects in the ensemble of donor electrons (green); we
 must additionally sample at the same values of $\tau$ as the 
experiment (red) yielding a stroboscopic effect.
(c) Comparison with experiment (black).
In our calculation, we combine decoherence effects of ESEEM, 
(non-Markovian) nuclear-induced SD, and
(Markovian) exponential relaxation 
by simply multiplying them together.  The orange
curve gives ESEEM of our theory [red curve in (b)] multiplied by
$\exp{\left[-2 \tau / T_R\right]} \exp{\left[-\left(2 \tau / T_{SD}\right)^n
\right]}$, where $T_R$, $T_{SD}$, and $n$ are fitting parameters for the
relaxation time, SD time, and SD exponent, respectively.
\label{AHF0deg}
}
\end{figure}

\begin{figure}
\includegraphics[width=3in]{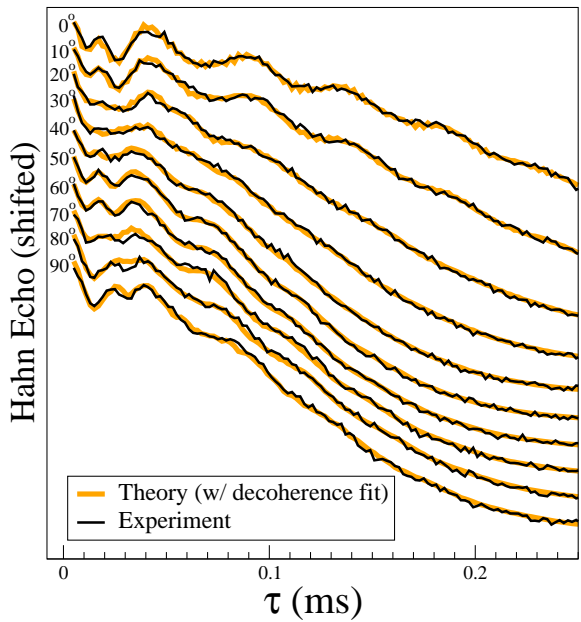}
\caption{
(Color online)
AHF-induced ESEEM in Si:P for ten different curves corresponding
  to ten different magnetic field angles ranging from the [001] to the
  [110] directions.  The plots are shifted in order to distinguish
  each angle.
All fits use the same normalization, strain
  distribution width ($0.4\%$), and relaxation time ($T_R = 2.17 \pm
  0.02~\mbox{ms}$) parameters.  
There are two fitting parameters per
  curve: the SD time, $T_{SD}$, and the SD exponent, $n$.  
These fitting parameters are compared with
  our SD theory\cite{witzelSD} in Fig.~\ref{AllAngleFits}.
\label{AllAngleAHFcomparison}
}
\end{figure}

\begin{figure}
\begin{center}
\includegraphics[width=3in]{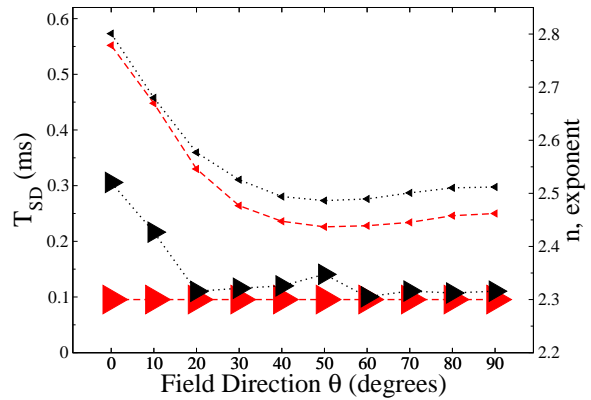}
\end{center}
\caption{
\label{AllAngleFits}
(Color online)
Comparison between the SD fitting parameters 
(black triangles connected with dotted lines)
of
Fig.~\ref{AllAngleAHFcomparison} and the theoretical predictions 
(red triangles connected with dashed lines)
of our SD\cite{witzelSD} theory.
Right [left] triangles correspond to
$n$ [$T_{SD}$]; sizes approximate fitting uncertainty.  
The experimental $n$ fit deviates from theory $(2.30
\pm 0.05)$ only at small angles, where 
nearest neighbor dipolar flip-flop interactions approach zero.
}
\end{figure}

Figure~\ref{AHF0deg}(c) shows excellent agreement of our ESEEM calculations
with experimental data reported in Ref.~[\onlinecite{TyryshkinSiCoherence}].  The theory calculations do use five separate fitting
parameters: normalization, strain distribution width, relaxation time, SD
time, and a SD exponent.  The first three of these parameters (described
momentarily) may be fixed for all different directions of the applied
magnetic field.  Thus, in Fig.~\ref{AllAngleAHFcomparison}, which shows
comparison with experiment\cite{TyryshkinSiCoherence} for ten different magnetic field directions, we
use only two fitting parameters per curve.  These two fitting parameters
characterize the SD decay and are compared with the results of our SD
theory\cite{witzelSD} in Fig.~\ref{AllAngleFits}.  Considering that 
we use, in our SD analysis, the Kohn-Luttinger envelope function
within the effective mass approximation for the
donor electron,\cite{Kohn-Luttinger} which is known
to be less reliable near the phosphorus atom,\cite{Kohn,Pantelides} the
agreement is quite good.  There appears to be, however, some discrepancy
between the fit and the theory for the SD exponent when the applied field is
close along the [001] lattice direction: theory expects $n=2.3$ and the
fit yields $n=2.5$.  It is probably not coincidental that the nearest
neighbor dipolar coupling vanishes when the applied field points along the
[001] direction.  Perhaps we have overlooked some interactions that become
important when the dipolar interaction is weak.  Overall, however, we do
demonstrate good theoretical understanding of Si:P electron spin decoherence
in Figs.~\ref{AHF0deg}, \ref{AllAngleAHFcomparison}, and \ref{AllAngleFits}.

We now describe the fitting parameters in more detail.  The AHF-induced ESEEM
[Eq.~(\ref{echo_mod})] for a single electron spin, using the experimentally
determined coupling constants for $22$ nuclear shells from
Ref.~[\onlinecite{Ivey75}] is shown in Fig.~\ref{AHF0deg}(a).  Random defects
such as dislocations cause strain effects (such as electron population shifts
between different Si valleys) that slightly alter the coupling constants for
different donors.  Strain effects result in narrow distributions for the
values of HF coupling constants and/or Zeeman frequencies and effectively
dampen the ESEEM signal for an ensemble of spins.  This is shown in
Fig.~\ref{AHF0deg}(b) where, in order to fit the experimental results,
we assume a Gaussian distribution for HF frequencies with a $0.4\%$
width.\cite{StrainZeeman}  We believe strain is the culprit of this
distribution of HF interaction strengths because of two reasons.  First, 
experimentalists observe strain effects orders of magnitude larger than this
when applying a small stress on their sample, so it is quite plausible and
likely that random defects are generating the strain, consistent with our
excellent fit.\cite{TyryshkinP,Hirayama,Yusa}  Second, other neglected
interactions, such as electrons interacting with nuclei at different donors
and dipolar interaction between nuclear spins, are much too small to account
for this effect.  If we do not assume a distribution of HF coupling
constants, the echo modulation would continue with a constant amplitude [see
Eq.~(\ref{echo_n})] until eventually nuclear spin relaxation kicks in.  This
is obviously not what is observed
experimentally.\cite{Tyryshkin60ms,TyryshkinSiCoherence,Abe,Ferretti}

In addition to strain, Fig.~\ref{AHF0deg}(b) shows the stroboscopic effect
that emerges when we sample the same values of $\tau$ as those reported in the
experiment.
The theoretical (orange) curve in Fig.~\ref{AHF0deg}(c) shows the
ESEEM result of Fig.~\ref{AHF0deg}(b) multiplied by $\exp{\left[-2 \tau /
T_R\right]} \exp{\left[-\left(2 \tau / T_{SD}\right)^n\right]}$ to account
for independent effects of Markovian relaxation and non-Markovian
nuclear-induced SD. 
After we normalize the signal strength as an
additional fit (the experiment only gives the Hahn echo decay on a relative
scale based on the strength of the observed signal),
we obtain excellent
agreement with the experimental results [black curve in
Fig.~\ref{AHF0deg}(c)].
Again, we use a total of five fitting parameters in
Fig.~\ref{AHF0deg}; however, we use only two fitting parameters
per curve in Fig.~\ref{AllAngleAHFcomparison} and these
two SD parameters are compared with theoretical
results via Ref.[\onlinecite{witzelSD}] in Fig.~\ref{AllAngleFits}.

Although nuclear-induced SD and ESEEM both result from interactions
with the nuclear spin bath, we have confirmed that it is appropriate to
treat them independently.  Nuclear-induced SD results from very small
contributions from many thousands of pairs of flip-flopping 
nuclei.\cite{witzelSD}
The weak nuclear dipolar coupling that is responsible for SD is too
small to have any significant impact on the ESEEM-contributing
nuclei that have a much stronger HF coupling to the electron.
To be rigorous, we have made explicit cluster expansion
calculations\cite{witzelSD} that include the AHF interaction.
ESEEM emerges from one-cluster contributions, and the two-cluster
contributions are negligibly
changed with the introduction of AHF coupling.  This calculation 
confirms the independence of these two decoherence channels.

The Markovian relaxation is dominated by instantaneous diffusion (ID)
that results from interactions between the resonant electron spin
donors.\cite{Schweiger}  Having a concentration of $8\times10^{14}$ 
donors / $\mbox{cm}^3$ in
these experiments,\cite{TyryshkinSiCoherence} with half of those at resonance with
the applied pulses
(determined by whether the P donor nucleus has an up or down spin), yields an
ID time of $T_{ID} \approx 3~\mbox{ms}$,\cite{Schweiger} accounting for
most of the $T_{R} \sim 2.2~\mbox{ms}$ relaxation.
A further relaxation process with a decay time of about $8.3~\mbox{ms}$
would be needed to account for the remaining contribution to $T_{R}$
(the inverse of the decay times are additive).
With these experiments performed at $8~\mbox{K}$, \cite{TyryshkinSiCoherence}
this may be attributed to temperature-dependent spin-lattice
relaxation; $8.3~\mbox{ms}$ is fairly consistent with 
reported
 $10-25~\mbox{ms}$ temperature-dependent relaxation time in Si:P at 
$8~\mbox{K}$ (a wide uncertainty range is due to strong temperature
dependence and offsets in temperature calibration).\cite{Tyryshkin60ms}

It is important to emphasize here that the theoretical fitting for the
echo modulations (separate from spectral diffusion and relaxation) uses 
experimentally obtained AHF interaction strength.  The only variable fitting
parameter for echo modulations is the $0.4\%$ distribution of the AHF
interaction strength, which allows us to reproduce the decay of the echo
modulation.  In the meantime, as indicated in Fig.~\ref{AllAngleFits}, the
fit for spectral diffusion is an attempt to use a relatively simple
functional form to represent our theoretical calculations in
Ref.~[\onlinecite{witzelSD}].  Therefore the seemingly large number of
fitting parameters involved here is not a blind attempt to obtain the best
possible fit to the experimental measurements.  Instead, they mostly provide
a more transparent representation of a more sophisticated theory, and the
remarkable agreement we obtain here illustrates that all the main decoherence
mechanisms are represented, most likely, in our description.


\section{Suppressing Anisotropic-Hyperfine-induced ESEEM}
\label{suppressingESEEM}

After building confidence in our theoretical approach by comparing our results
with experiments, we now address the question of how to suppress spin
decoherence induced by AHF interaction.  More specifically, we will show that
this AHF-induced decoherence may be drastically reduced by carefully choosing
the strength and direction of the applied magnetic field and by applying spin
echo pulses with precisely prescribed timing.

One interesting and important feature of the maximum modulations shown in
Fig.~\ref{FigModDepth}~(a) is that a peak occurs when $\omega_I \sim A_n/2$
(with $A_n$ positive) for each shell of atoms.  At each such cancellation
condition, as it is dubbed, the Zeeman and HF energies of nucleus $n$ cancel
when the electron spin is up but not down, freeing the nuclear spin from
conservation of energy constraints conditional upon the state of the electron
spin.  Mathematically, $\omega_{n+}$ is minimized [Eq.~(\ref{omega_n})] so
that $k_n$ is at (or very near) its maximum, resulting in modulation depth
peaks.  This effect is shown experimentally by comparing the two echo decay
curves in Fig.~\ref{FigModDepth}~(c); the curve corresponding to a field
strength closer to the center of the A-shell peak clearly exhibits stronger
echo modulations.  It turns out that the experiments discussed in
Sec.~\ref{vsExperiment} were performed with a magnetic field strength denoted
as $\chi$ in Fig.~\ref{FigModDepth}~(b), which is very close to the center of
the A-shell peak, where echo modulations are particularly strong.

\begin{figure}
\includegraphics[width=3in]{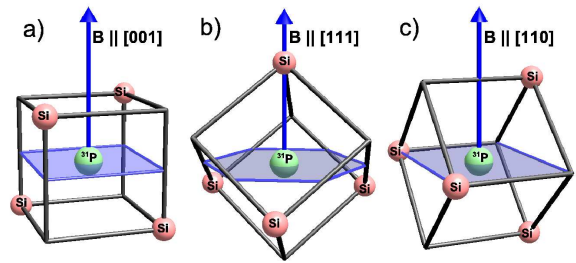}
\caption{
(Color online)
Special applied magnetic field directions that allow 
effective removal of echo modulation contributions due to E-shell
nuclei (the four nearest neighbors to the P donor).
The arrows and translucent sheets, respectively, 
indicate directions from the P atom 
parallel and perpendicular to the applied magnetic field.
Sites in these direction 
give no anisotropic contribution ($B_n = 0$) because these directions
are, by symmetry, along the principal axes of the ${\bf A}$ tensor of
Eq.~(\ref{HFtensor}); 
thus, the
top site in (b) and the two in-plane sites in (c) do not contribute to ESEEM.
In each of the three cases, the E-shell sites that {\it do} contribute
are magnetically equivalent.   
\label{SpecialOrientations}
}
\end{figure}

It is clear from this discussion that to minimize AHF-induced decoherence,
cancellation conditions for all the shells with finite AHF coupling constant
should be carefully avoided by properly selecting the applied magnetic field
strength (or, in electron spin resonance, the corresponding microwave cavity
frequency).  Furthermore, away from the cancellation condition peaks, the
E-shell nuclei (nearest neighbors to the P nucleus) have the strongest AHF
coupling by far, so that they dominate the echo modulations by more than an
order of magnitude, as seen in Fig.~\ref{FigModDepth}~(a).  Remarkably, the
echo modulation due to these dominating E-shell nuclei can be effectively
removed at special magnetic field orientations
(Fig.~\ref{SpecialOrientations}), if we exploit the periodic restoration of
electron spin coherence in the presence of nuclei that are seen by the
electron as magnetically equivalent.  This restoration arises because
$V_n(\tau) = 1$ [Eq.~(\ref{echo_n})] when $\tau$ is a multiple of $2
\pi / \omega_{n\pm}$ (either $+$~or~$-$).  Note that such periodic
restoration does {\it not} generally occur in the free induction case
of Sec.~\ref{FID}.
%
%

\begin{figure}
\includegraphics[width=3in]{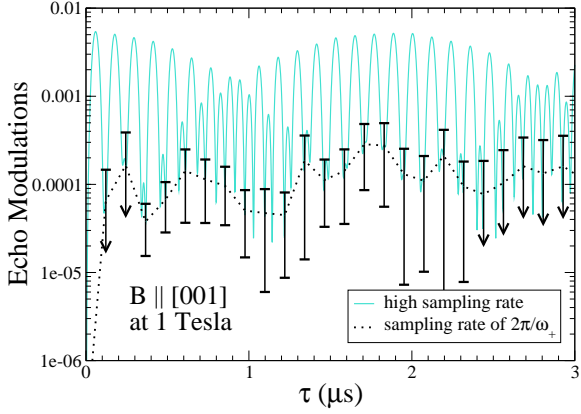}
\caption{
(Color online) Echo modulations, $1 - V(\tau)$, in natural Si with an applied
field of $1~\mbox{T}$ in the $[001]$ direction corresponding to
Fig.~\ref{SpecialOrientations}~(a).  When sampling $\tau$ at multiples of $2
\pi / \omega_+$ (or $2 \pi / \omega_-$), the E-shell nuclei give no
contribution to the echo modulations.
%
%
Error bars, vertically asymmetric
because of the logarithmic scale, 
correspond to the standard deviation resulting from random isotopic
configurations; those with down arrows 
extend below the visible range.
\label{FigPeriodicRestoration}
}
\end{figure}

For special magnetic field orientations shown in
Fig.~\ref{SpecialOrientations}, the contributing E-shell sites are
magnetically equivalent with the same $\{\omega_{n-}, \omega_{n+}\}$; thus,
the electron spin is periodically restored at the same values of $\tau$
regardless of isotopic ($^{29}$Si) configuration.  In this way, E-shell
contributions can effectively be eliminated as exemplified in
Fig.~\ref{FigPeriodicRestoration}.  By orienting the magnetic field in one of
the various special directions, the effects of all E-shell nuclei are
simultaneously eliminated at periodic values of $\tau$.

To understand the periodic restoration of ESEEM in the presence of
magnetically equivalent nuclei, note that $\hat{U}_0^{\pm}(t)$ simply
generates precession for each nucleus at a frequency of $\omega_{n\pm}$,
respectively.  For instance, given an electron spin that is up, the nuclear
spin returns to its initial orientation after waiting for a time that is a
multiple of $2 \pi / \omega_{+}$.  That is, assuming all nuclei being
considered are magnetically equivalent, $\hat{U}_0^{\pm}(2 \pi m /
\omega_{\pm}) = \identity$ for any integer $m$.  If we consider a Hahn echo
sequence with $\tau = 2 \pi / \omega_{+}$, we have [see
Eq.~(\ref{HahnEvolUpDown})] $\hat{U}_{\mbox{\tiny Hahn}}^{\pm} =
\hat{U}^{\mp}_0(\tau) \hat{U}^{\pm}_0(\tau) = \hat{U}_0^{-}(\tau)$.  Thus the
evolution of the magnetically equivalent nuclei is independent of the electron
spin, since $\hat{U}_{\mbox{\tiny Hahn}}^{+} = \hat{U}_{\mbox{\tiny
Hahn}}^{-}$, so that the electron qubit is fully decoupled from these nuclei
(apart from the effects of the neglected nuclei that are not magnetically
equivalent to the others).  This property is common to any balanced sequence
in which an initially up or down electron (or the separate components of a
superposition state) spends an equal amount of time being up and down. 
Specifically, we may apply our timing trick to eliminate E-shell modulations
for the Carr-Purcell sequences\cite{Meiboom, witzelCPMG} or concatenated
dynamical decoupling sequences\cite{KhodjastehPRL, yaoCDD} that can be much
more effective than the Hahn echo for mitigating other types of decoherence
such as spectral diffusion.\cite{witzelCPMG, yaoCDD}

It is important to determine how sensitive our technique of suppressing 
AHF-induced decoherence is to errors in timing and magnetic field direction.
Because the E-shell contributions oscillate at frequencies of $\omega_{n-}$
and $\omega_{n+}$, the timing of the pulse sequence in order to remove this
effect must be accurate with errors that are small compared with the period
of these oscillations, $2 \pi / \omega_{n-}$ and $2 \pi / \omega_{n+}$.  For
magnetic fields that are $\gtrsim~1~\mbox{T}$, these oscillation frequencies
[Eq.~(\ref{omega_n})] are dominated by the nuclear Zeeman frequency,
$\omega_{n\pm} \approx \omega_I$, so that the periods are approximately $2
\pi / \omega_I \sim 100~\mbox{ns}$, as is apparent in
Fig.~\ref{FigPeriodicRestoration}.  

The timing is also affected by systematic uncertainty in either $\omega_{n+}$
or $\omega_{n-}$ (whichever we choose for synchronization) in a way that
worsens in time and number of oscillation periods.  The resulting uncertainty
in time is given by
\begin{equation}
\delta \tau = \left(\frac{\delta \omega_{n\pm}}{\omega_{n\pm}}\right)
\tau.
\end{equation}
Then, the requirement that $\delta \tau \ll 2 \pi / \omega_{n\pm}$ yields the
following restriction:
\begin{equation}
\tau \ll \frac{2 \pi}{\delta \omega_{n\pm}}.
\end{equation}
When dealing with an ensemble, the uncertainty of $\omega_{n\pm}$ will be
subject to strain effects.  In Sec.~\ref{vsExperiment}, we empirically
determined strain effects causing a $0.4\%$ distribution, which equates to a
frequency uncertainty of about $10~\mbox{kHZ}$.  This implies that we are
limited to $\tau \ll 100~\mbox{$\mu$s}$.  On the other hand, the situation
in a different experiment or sample could be quite different and would depend
on the precise effects of strain on the E-shell nuclei which is outside of
the scope of this paper.  Also, depending on the architecture of the proposed
quantum computer, it may be possible to calibrate pulse timing to individual
qubits rather than an ensemble, circumventing the uncertainty caused by
strain.

Errors in the applied magnetic field direction {\it cannot} be fixed with
calibration.  If the magnetic field direction deviates from one of the three
special directions in Fig.~\ref{SpecialOrientations}, the contributing
E-shell nuclei will no longer be equivalent and nuclei that are supposed to
be inert [e.g., the top nucleus in Fig.~\ref{SpecialOrientations}~(b) and the
two leftmost nuclei in Fig.~\ref{SpecialOrientations}~(c)] will, in general,
contribute to ESEEM.  In order to address the question of sensitivity to the
magnetic field direction, we first consider how angular uncertainty affects
the uncertainty of $\omega_{n\pm}$:
\begin{equation}
\nonumber
\frac{\delta \omega_{n\pm}}{\delta \theta} = \frac{1}{\omega_{n\pm}}
\left[\frac{\pm 1}{2} \left(\pm A_n/2 - \omega_I\right) 
\frac{\delta A_n}{\delta \theta}
+ \frac{1}{4} B_n \frac{\delta B_n}{\delta \theta} \right].
\end{equation}
We again assume that $\omega_{n\pm} \sim \omega_I$, so that
\begin{equation}
\max_n \left \lvert \frac{\delta \omega_{n\pm} }{\delta \theta} 
\right\rvert
\approx \frac{1}{2} \max_n \left\lvert \frac{\delta A_n}{\delta \theta} 
\right\rvert.
\end{equation}
This is valid when $\max_n \lvert A_n \rvert \gtrsim \max_n \lvert B_n
\rvert$, so that the $\delta B_n$ contribution is negligible.  We have
verified this numerically by calculating how $A_n$ and $B_n$ vary with
respect to changing the magnetic field direction about an axis that rotates
from $[100]$ to $[011]$.  Although we did not analyze full two-dimensional
changes in direction, we do extract a characteristic scale of 
$50~\mbox{kHz}/\mbox{deg}$ for both $\delta A_n / \delta \theta$ and
$\delta B_n / \delta \theta$ of the E-shell nuclei, which we expect to be
general.  In addition to $\omega_{n\pm}$, we must also consider how an error
in the magnetic field direction away from $[111]$ or $[011]$ affects $k_n$ for
those nuclei which are not expected, ideally, to contribute.  For such
nuclei,
\begin{eqnarray}
k_n &\approx& \left. \frac{1}{2} \frac{\delta^2 k_n}{\delta \theta^2}
\right\rvert_{B_n = 0} \left(\delta \theta\right)^2, \\
\nonumber
\left. \frac{\delta^2 k_n}{\delta \theta^2}
\right\rvert_{B_n = 0} &=&  
\frac{2 \omega_I^2}{\left(\omega_{n+} \omega_{n-}\right)^2} 
\left(\frac{\delta B_n}{\delta \theta} \right)^2 \\
&\approx& 2 \left(\frac{\delta B_n / \delta \theta}{\omega_I} \right)^2.
\end{eqnarray}
We find that $k_n B / (\delta \theta)^2 \sim 0.5\%~\mbox{T}^2 / \
\mbox{deg}^2$.  Thus, with a one degree error in the magnetic field
direction and a field strength of $1~\mbox{T}$, the sites that are not
supposed to contribute to ESEEM when the field is oriented along $[100]$ or
$[011]$ could, if occupied by a $^{29}$Si nucleus, actually provide
modulations up to about $1\%$.  This, however, scales quadratically with the
angle error, so an error of $1\%$ of a degree would keep these errors below
the $10^{-6}$ threshold often quoted for quantum error correction.


\section{Conclusion}
\label{conclusion}

We have studied Si:P donor electron spin decoherence due to AHF interaction,
which is an important dephasing mechanism in Si.  We clarify the electron spin
echo envelope modulation in the Si:P system and the resonance-like
contributions from nuclear spins in various shells away from the P atoms. 
Our theory is in excellent quantitative agreement with experiment.  Most
importantly, we suggest an approach to minimize the decoherence effect of AHF 
interaction by avoiding the cancellation conditions and orienting an applied
magnetic field along directions that can periodically eliminate the
contributions from the dominant E-shell nuclei.  We quantitatively analyze
the sensitivity of this strategy to errors in timing and magnetic field
orientation.  

In principle, one can eliminate the problem of nuclear spin decoherence
through isotopic purification, removing the $^{29}$Si.  After purification,
those donors that still have $^{29}$Si nuclei in the E-shell may be
disqualified as qubits.  Our strategy provides an alternative solution to
this problem that provides the quantum computer architect with more
flexibility.

We wish to thank David Cory, Steve Lyon, Sergei Dikanov, and Semion Saikin for
useful discussions.  We especially would like to thank Alexei Tyryshkin for
his invaluable insight, suggestions, and experimental data.  We thank NSA,
LPS, and ARO for support.


\begin{thebibliography}{99}
\bibitem{StrainZeeman}Similar results are obtained if the Zeeman 
frequencies are randomly distributed by $0.4\%$ instead of the 
HF frequencies, corresponding to a random field variation in the
range of 10 Gauss, (or 
in some combination such that they combine in quadrature to $0.4\%$),
though we believe that strain is more likely to affect
HF interactions via small distortions in the electron wave function.
\bibitem{Hirayama} Small strains can result even from mundane sources
in an experiment.  For example, in a private conversation, Y. Hirayama
of the
University of Tohoku pointed out that in a NMR experiment on a GaAs quantum 
point contact\cite{Yusa}, strain from the way the sample
is mounted can result in a 15 kHz nuclear quadrupole splitting.
\end{thebibliography}
\end{document}